\documentclass[journal]{IEEEtran}

\usepackage{multirow}
\usepackage{longtable}
\usepackage{amsmath,amssymb,amsfonts}
\usepackage{algorithmic}
\usepackage{graphicx}
\usepackage{comment}
\usepackage{textcomp}
\usepackage{xcolor}
\usepackage{color}
\usepackage[inline]{enumitem}
\usepackage[ruled,vlined,linesnumbered]{algorithm2e}
\usepackage{lscape}
\ifCLASSINFOpdf
\else
\fi
\begin{document}
%
\title{A Survey on  Secure and Private Federated Learning Using Blockchain: Theory and Application in Resource-constrained Computing}
%
%
%
\author{Ervin~Moore,
        ~Ahmed~Imteaj,
        ~Shabnam~Rezapour,
        ~and
        ~M.~Hadi~Amini,~\IEEEmembership{{\color{black}Senior Member,~IEEE}}
\thanks{Corresponding Author: M. Hadi Amini, Florida International University, Miami, FL 33199, moamini@fiu.edu}

\thanks{Ervin Moore and M. Hadi Amini are with Knight Foundation School of Computing and Information Sciences, Florida International University, Miami, FL 33199, USA. They are also with the Sustainability, Optimization, and Learning for InterDependent networks laboratory (solid lab) at FIU.}
\thanks{Ahmed Imteaj is with Sustainability, Privacy and Edge intElligence for Distributed networks (SPEED) Laboratory, Southern Illinois University, Carbondale, USA.}
\thanks{Shabnam Rezapour is with Smart Decision-Making for Network-Centric Systems, Florida International University, Florida, USA.}

}

%
%

\markboth{}
{Moore \MakeLowercase{\textit{et al.}}: }
%



\maketitle

\begin{abstract}
Federated Learning (FL) has gained widespread popularity in recent years due to the fast booming of advanced machine learning and artificial intelligence along with emerging security and privacy threats. FL enables efficient model generation from local data storage of the edge devices without revealing the sensitive data to any entities. While this paradigm partly mitigates the privacy issues of users' sensitive data, the performance of the FL process can be threatened and reached a bottleneck due to the growing cyber threats and privacy violation techniques. To expedite the proliferation of FL process, the integration of blockchain for FL environments has drawn prolific attention from the people of academia and industry. Blockchain has the potential to prevent security and privacy threats with its decentralization, immutability, consensus, and transparency characteristic. However, if the blockchain mechanism requires costly computational resources, then the resource-constrained FL clients cannot be involved in the training. Considering that, this survey focuses on reviewing the challenges, solutions, and future directions for the successful deployment of blockchain in resource-constrained FL environments. We comprehensively review variant blockchain mechanisms that are suitable for FL process and discuss their trade-offs for a limited resource budget. Further, we extensively analyze the cyber threats that could be observed in a resource-constrained FL environment, and how blockchain can play a key role to block those cyber attacks. To this end, we highlight some potential solutions towards the coupling of blockchain and federated learning that can offer high levels of reliability, data privacy, and distributed computing performance. 
\end{abstract}

\begin{IEEEkeywords}
Federated Learning, Blockchain, Security, Privacy, Resource Limitations, Optimization.
\end{IEEEkeywords}

%
\IEEEpeerreviewmaketitle

\section{Introduction}

\subsection{Motivation, Comparison, and Contributions}
\IEEEPARstart{D}{}ata-driven technologies can be limited by factors such as limited computing resources or the need for large amounts of quality data. Earlier online methodologies transferred raw data which created potential data privacy risks. Data silos are formed in organizations that are inaccessible to other departments and often incompatible with datasets within the same organization. FL is introduced as a solution for privacy preservation due to FL applying differential privacy to communicated data. In FL machine learning parameters are exchanged instead of raw data. Centralized computing architectures can be directly targeted and denied because of the single-point-of-failure vulnerability. Blockchains being a decentralized approach can improve system resilience from adversaries. A combination of FL and blockchain have increased robustness in comparison to earlier methodologies that had diminishing performance, due to features such as lack of privacy preservation. This survey examines blockchain-based FL as a solution for privacy preservation and secure resource management. Recent blockchain and FL surveys did not comprehensively analyze combining the two to provide a secure learning setting for a resource-constrained environment. 

\vspace*{-0.25cm}

\subsection{Organization of the Survey}
{\color{black}
The rest of this paper is organized as follows. 
{\color{black}In Section \textbf{II}, we present an overview and taxonomy of FL with a comprehensive list of existing studies. In Section \textbf{III}, we review distributed optimization and ML approaches. Section \textbf{IV} presents a detailed analysis of the major challenges of FL while applying on resource-constrained devices, which is followed by Section \textbf{V}, where we discuss the potential solutions of those emerging challenges. In Section \textbf{VI}, we present the existing FL applications, and in Section \textbf{VII}, we highlight the future research direction in the FL-based IoT domain. Finally,  Section \textbf{VIII} concludes the paper.}
}



\section{Background Study}
{\color{black}
Online computing environments can put private information at risk when communicating online. Ensuring data privacy throughout communications is important and can be approached by FL. FL transmits user data through machine learning updates instead of raw data. Classical data-sharing techniques can be improved by FL and its privacy protection mechanisms. Blockchain, a decentralized approach, offers additional security and resources for applications such as FL. BCFL is an appropriate solution for resource-constrained computing environments due to: peer-to-peer computing, network participation incentives, validation protocols, and other resource-preserving mechanisms. The following section details mechanisms within FL and blockchain. 

}
\subsection{Brief Introduction to Federated Learning}
FL proposed by Google in 2016, protects users' privacy by allowing devices to train a machine learning model using local data collaboratively. FL can process model updates synchronously or asynchronously. FL is a collaborative machine-learning architecture that stores local data on a device, this local data reduces the need for data to be stored online in a cloud. Sensitive user data is protected by adding noise to personally identifiable information. The goal is for data to be unidentifiable while preserving good qualities for machine learning model optimization. Finding the balance of data retention improves performance. FL, as a manner of distributed machine learning, can significantly preserve clients’ private data from being exposed to external adversaries \cite{wei2020federated}.

Data structures encourage distinct FL implementations for optimal data management. Each FL structure looks at feature spaces and datasets differently. Three different types of FL implementations include Horizontal FL, Vertical FL, and Federated Transfer Learning. 
\begin{itemize}
    \item Horizontal (sample-based) FL - Data distribution contains a consistent set of features and different samples. For example, a database containing predefined rows of relevant user data has different entries for each unique user.
    Some examples of Horizontal FL are predicting smartphone user behavior, personalized recommendations, identifying risk factors or predicting patient diseases, optimizing manufacturing product outcomes, and recognizing fraudulent transactions \cite{suzen2020novel}.
 
    \item Vertical (feature-based) FL - Data distribution contains recurring samples with different features. For instance, consider a bank and a superstore in the same area. Most of their customers may be the same, but their business structure, i.e., the feature space, is different, and thus the user-space intersection is quite large \cite{imteaj2021survey}. Some examples of vertical FL are predicting the likelihood of patient admission considering different types of health records, lab results in hospitals, customer shopping behavior, and likelihood to make purchases analyzing heterogeneous data types of multiple retail stores \cite{yang2019federated}.

    \item Federated Transfer Learning (FTL) - Combination of horizontal and vertical FL that contains different features and samples. The knowledge of an existing machine learning model is transferred to another model for improved performance. Transfer learning reuses learned lessons and re-purposes information towards a related problem. FTL currently has applications in wearable healthcare \cite{chen2020fedhealth}, autonomous driving, classification of EEG signals \cite{raza2022designing}, industrial fault diagnostics \cite{zhang2021federated}, and image steganalysis \cite{yang2020fedsteg, saha2021federated}.
\end{itemize}

\begin{figure}[htp]
\begin{center}
    \includegraphics[width=1\linewidth]{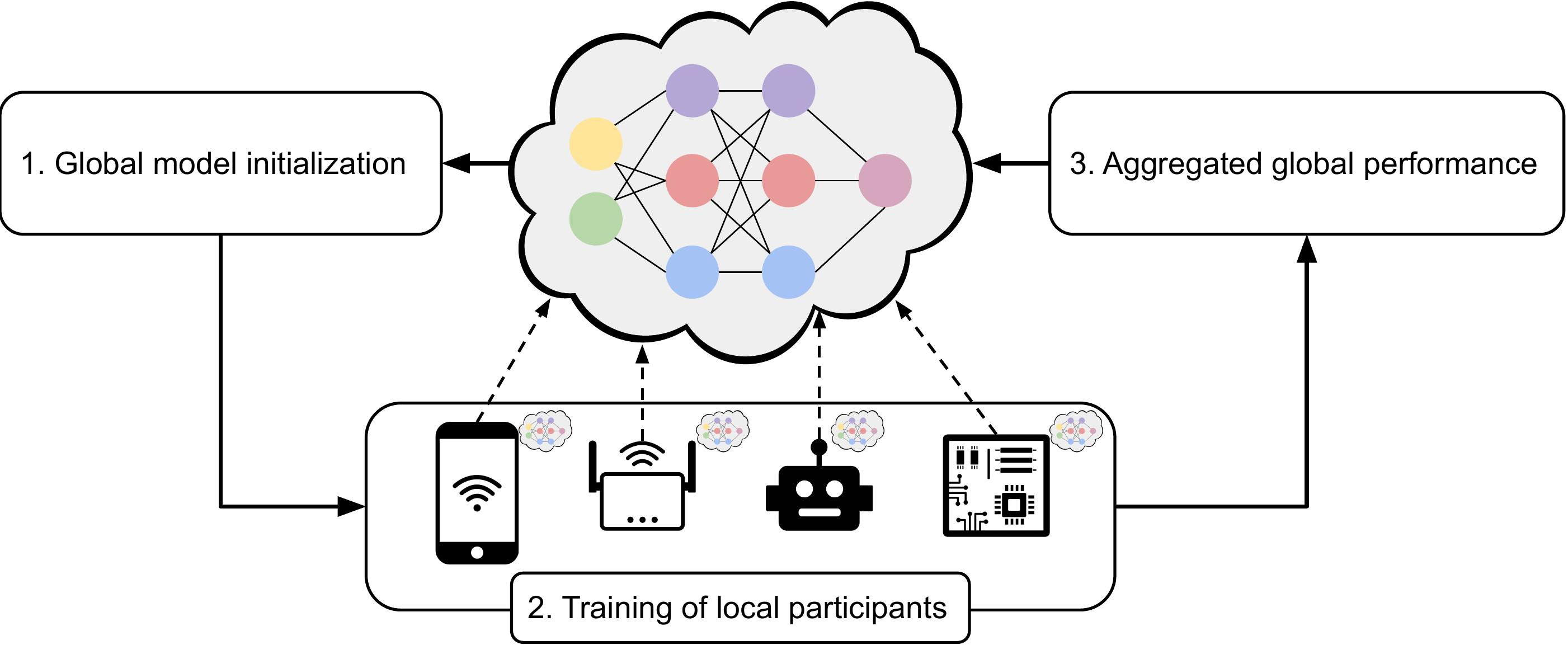}
    \caption{Introduction to FL cycle.}
    \label{fig:FL_diagram}
\end{center}
\end{figure}

Traditional FL has a single point of failure limitation. Figure \ref{fig:FL_diagram} showcases an uninterrupted online learning environment, although a denial of service could halt the repeated process. Internet of Things (IoT) devices differ in performance and reliability. The FL training process could be computationally expensive depending on available resources, which may require devices to drop out from the learning process. Devices may need the motivation to participate in FL training, which can be encouraged by incentives. Many of the pitfalls of FL can be augmented by blockchain mechanisms.

\subsection{Brief Introduction to Blockchain}

Blockchain technology connects data blocks into a digital ledger similar to a database. The blockchain ledger is distributed throughout the network and considered decentralized; thus, third-party intermediaries are not required for blockchain processes. Each block in a blockchain contains relevant transaction history, such as unique identifiers, block computing costs, and other network-related information. Network transactions are directly recorded into the blockchain once authenticated. Peer-to-peer transactions allow workloads to share resources and information. Peers can participate as suppliers and consumers of resources. The authors in \cite{imteaj2021foundations} proposed that the nodes or agents involved within a blockchain network are called participants and miners. The participants are the agents who perform any transaction, and the miners are responsible for validating or rejecting a block \cite{imteaj2021foundations}. In a blockchain, network processing happens simultaneously, creating a large optimization surface area. Resources such as processing power, storage space, and available network bandwidth all have to be considered for resource-constrained computing environments.

Many processes operate at the same time within a blockchain. When blocks are being generated at a high rate, performance can be lower. Merge issues can appear when two 

\begin{figure}[htp!]
    \centering
    \includegraphics[width=1\linewidth]{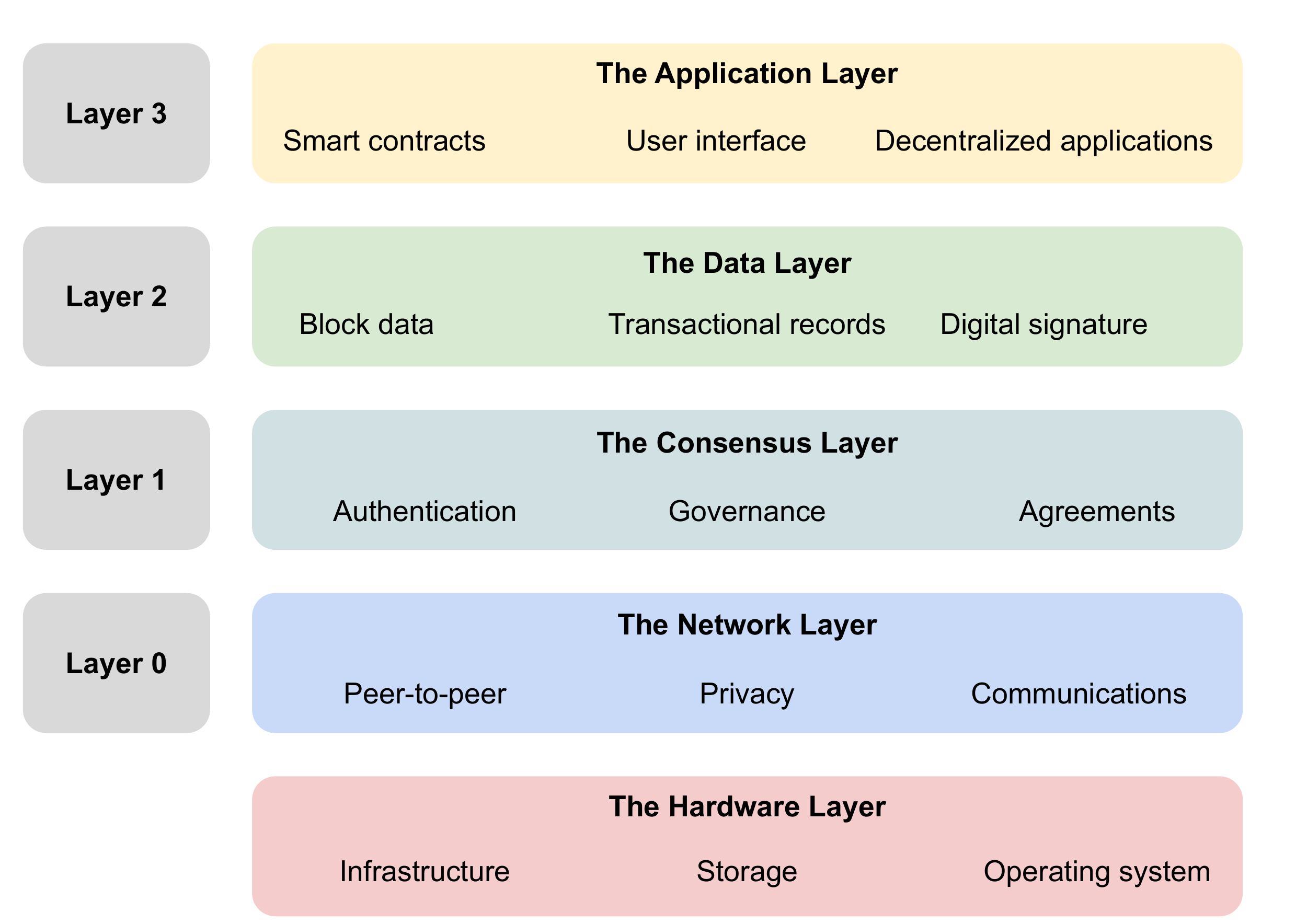}
    \caption{Blockchain architecture layers.}
    \label{fig:Blockchain_layers}
\end{figure}

\noindent or more blocks with the same hash information are successfully mined simultaneously. Merge issues can create forks in the blockchain, causing alternative chains to emerge until solved by consensus. Having a proper consensus protocol can discourage multiple versions of a blockchain. A rule of thumb is to reference the largest chain of blocks as the main blockchain to deter conflicts. Protecting the integrity of blockchain allows historical data to sync efficiently.

Blockchain architectures contain several layers:
\begin{itemize}
    \item The Hardware layer is necessary for hosting a blockchain architecture. The core infrastructure requires computers, graphics processing units (for miners), and miscellaneous data storage. Computing resources can be emulated through virtual machines as a process.
    \item The Network layer is responsible for peer-to-peer communication. Block transaction data is publicly visible in a public ledger. As a result, public blockchain architectures can share processes with unknown participants that can directly interact with one another.
    \item The Consensus layer authenticates transactions based on agreed-upon protocols. This layer is responsible for validating transactions. If a fork in the blockchain emerges, the consensus layer handles disputes using protocol logic. 
    \item The Data layer includes blocks and block transactional records. Blocks specifically contain information about: the previous block hash, the timestamp of the block creation, the generated block hash, and the transaction nonce, which is used for verifying transactions.
    \item The Application layer contains the front end of the blockchain and is considered the visible layer for the user. This layer contains an application programming interface (API), displays smart contracts, and includes basic configuration for quickly exchanging information with other blockchains. Blockchain interoperability allows similar blockchain structures to communicate 
    efficiently. 
\end{itemize}

\subsubsection{Blockchain Fundamentals}
\hfill\\
Blockchain architectures use cryptographic hash functions for transactions. Miners solve cryptographic puzzles to generate blocks and may form mining pools to boost performance. Nonce (numbers only used once) are numbers that miners must calculate before solving a cryptographic block. Cryptographic hash functions are encrypted, requiring efficient computing techniques to reasonably solve. The consensus layer authenticates that block conditions are met before integrating blocks into the blockchain. Blockchain agreements are transacted through smart contracts, rule-based digital agreements between two parties (i.e., miners and participants) that automatically execute when conditions are met. Once a block is authenticated, a new block is sought for mining. The authors in \cite{imteaj2021foundations} proposed that typically in a blockchain environment, constant creation of new blocks, even with no transaction, is crucial for maintaining security as it prevents malevolent users from creating longer, tampered blockchain \cite{imteaj2021foundations}.
\subsubsection{Blockchain Categories}
\hfill\\
Generally, there are three types of blockchains: public, private, and consortium. Each type of blockchain involves different permissions and participation specifics. Different types of consensus and authority mechanisms are included in each blockchain type.
\begin{itemize}
    \item Public blockchain - The most well-known blockchain is the public blockchain which is permissionless and open to all participants. Public blockchain ledgers are publicly available. Each participant has equal permissions in this peer-to-peer networking environment. Open participation in blockchains has been known to affect the number of resources in many ways. Cases, where open participation leads to a large number of dishonest participants can cause blockchain performance to diminish. Examples of popular public blockchains include cryptocurrency platforms such as Etheruem or Bitcoin. 
    \item Private blockchain - Private blockchains are commonly referred to as permissioned blockchains. Only authorized participants are allowed to join and view private blockchain ledgers. The central entity of the private blockchain has permission to alter and manage protocols. The authors in \cite{imteaj2021foundations} mention that the central entity has the power of validation and can change any rule of the blockchain (e.g., block consensus) \cite{imteaj2021foundations}.
    \item Consortium blockchain - Consortium blockchains include public and private blockchain qualities. Consortium blockchains are partially decentralized, permissioned and transparent. Selected authorities have increased permissions compared to the public. Increased permissions for consortium authorities include the ability to participate in block validation. The authors in \cite{qi2021privacy} utilized a consortium blockchain for pre-selecting a limited number of trusted miners to maintain the distributed ledger of an intelligent transportation system.
\end{itemize}

\begin{table}[!htp]
\tiny
\centering
\caption{\label{comparison-table}Summary of included papers regarding FL, Blockchain and Blockchained FL (BCFL).}
\setlength\tabcolsep{3.5pt}
\begin{tabular}{|l|l|l|l|l|l|} 
\hline
\begin{tabular}[c]{@{}l@{}}Ref\\ No.\end{tabular} & Area & Year & Contribution & \begin{tabular}[c]{@{}l@{}}Resource-\\constrained \\ considerate \end{tabular} & \begin{tabular}[c]{@{}l@{}}Participation \\incentivized \end{tabular} \\ \hline
\cite{baccarelli2022afafed} & FL & 2022 & \begin{tabular}[c]{@{}l@{}}Adaptive FL survey for minimizing resource consumption \\ within communication constrained environments \end{tabular} &  Yes & No\\ \hline
\cite{blanco2021achieving} & FL & 2021 & \begin{tabular}[c]{@{}l@{}}
Comprehensive FL survey focusing on security and privacy in FL \\ systems, the authors examine FL limitations and countermeasures \end{tabular} &  No & No\\ \hline
\cite{cheng2022blockchain} & BCFL & 2022 & \begin{tabular}[c]{@{}l@{}} Blockchain empowered FL model for blade icing estimation \end{tabular} &  No & Yes \\ \hline
\cite{cui2022fast} & BCFL & 2022 & \begin{tabular}[c]{@{}l@{}} Article that proposes a Blockchain-based communication-efficient \\ federated learning framework that compresses communications\end{tabular} &  Yes & No \\ \hline
\cite{feng2021blockchain} & BCFL & 2021 & \begin{tabular}[c]{@{}l@{}} Blockchain-empowered decentralized horizontal FL framework \\ that improves 5G-Enabled UAV privacy \end{tabular} &  Yes & No \\ \hline
\cite{imteaj2021foundations} & Blockchain & 2021 & Comprehensive book on fundamental topics related to Blockchain & No & Yes \\ \hline
\cite{imteaj2021survey} & FL & 2021 & \begin{tabular}[c]{@{}l@{}}Comprehensive survey on FL challenges that occur when \\ applied to resource-constrained Internet of Things devices\end{tabular} & Yes & Yes \\ \hline
\cite{issa2022blockchain} & BCFL & 2022 & \begin{tabular}[c]{@{}l@{}}Comprehensive Blockchained FL survey regarding approaches \\ for securing Internet of Things devices\end{tabular} & Yes & Yes\\ \hline
\cite{khor2021public} & Blockchain & 2021 & \begin{tabular}[c]{@{}l@{}}Comprehensive survey on public Blockchain for Internet of \\ Things devices with bounded computing capabilities\end{tabular} & Yes & Yes\\ \hline
\cite{kim2019blockchained} & BCFL & 2019 & Blockchained FL paper that examines end-to-end latency & No & No\\ \hline
\cite{li2022blockchain} & BCFL & 2022 & \begin{tabular}[c]{@{}l@{}}In-depth survey on Blockchained FL as a framework \end{tabular} & No & Yes\\ \hline
\cite{nguyen2021federated} & BCFL & 2021 & \begin{tabular}[c]{@{}l@{}}Blockchained FL paper that explores opportunities and \\ challenges in multi-access edge computing\end{tabular} & Yes & Yes\\ \hline
\cite{nguyen2022federated} & FL & 2022 & Survey on FL applications within smart healthcare & Yes & Yes\\ \hline
\cite{truex2019hybrid} & FL & 2019 & \begin{tabular}[c]{@{}l@{}} Hybrid FL system that protects against inference threats and \\ produces models with high accuracy.\end{tabular} & No & No\\ \hline
\cite{wang2019adaptive} & FL & 2019 & \begin{tabular}[c]{@{}l@{}}Adaptive FL paper focused on gradient-descent based FL for \\ reducing computing resource budgets\end{tabular} & Yes & No\\ \hline
\cite{yang2019federated} & FL & 2019 & \begin{tabular}[c]{@{}l@{}}Comprehensive survey on FL concepts and applications \end{tabular} & No & Yes \\ \hline
\cite{zhao2020privacy} & BCFL & 2020 & \begin{tabular}[c]{@{}l@{}}Privacy-preserving FL paper that leverages customers data \\ through IoT devices to assist home appliance manufacturers\end{tabular} & Yes & Yes\\ \hline
\end{tabular}
\end{table}

Consortium or permissioned blockchains are suitable when the integrity of participants is in question. Dishonest participants may sabotage others to compete for incentives or resources. Table \ref{comparison-table} summarizes various papers that consider participation incentives and resource constraints. The following section examines potential attacks, vulnerabilities, and threats found in BCFL environments.

\subsection{Threat Models and Integration Motivation}
FL includes privacy protection mechanisms that deter unwanted access to data. Meticulous attack attempts can breach FL security in different ways. Attacks can occur during device training, parameter upload and download, central aggregation, and post-aggregation phases. A range of various attacks can exploit FL vulnerabilities during each stage:

\begin{figure}[htp!]
    \centering
    \includegraphics[width=1.2\linewidth]{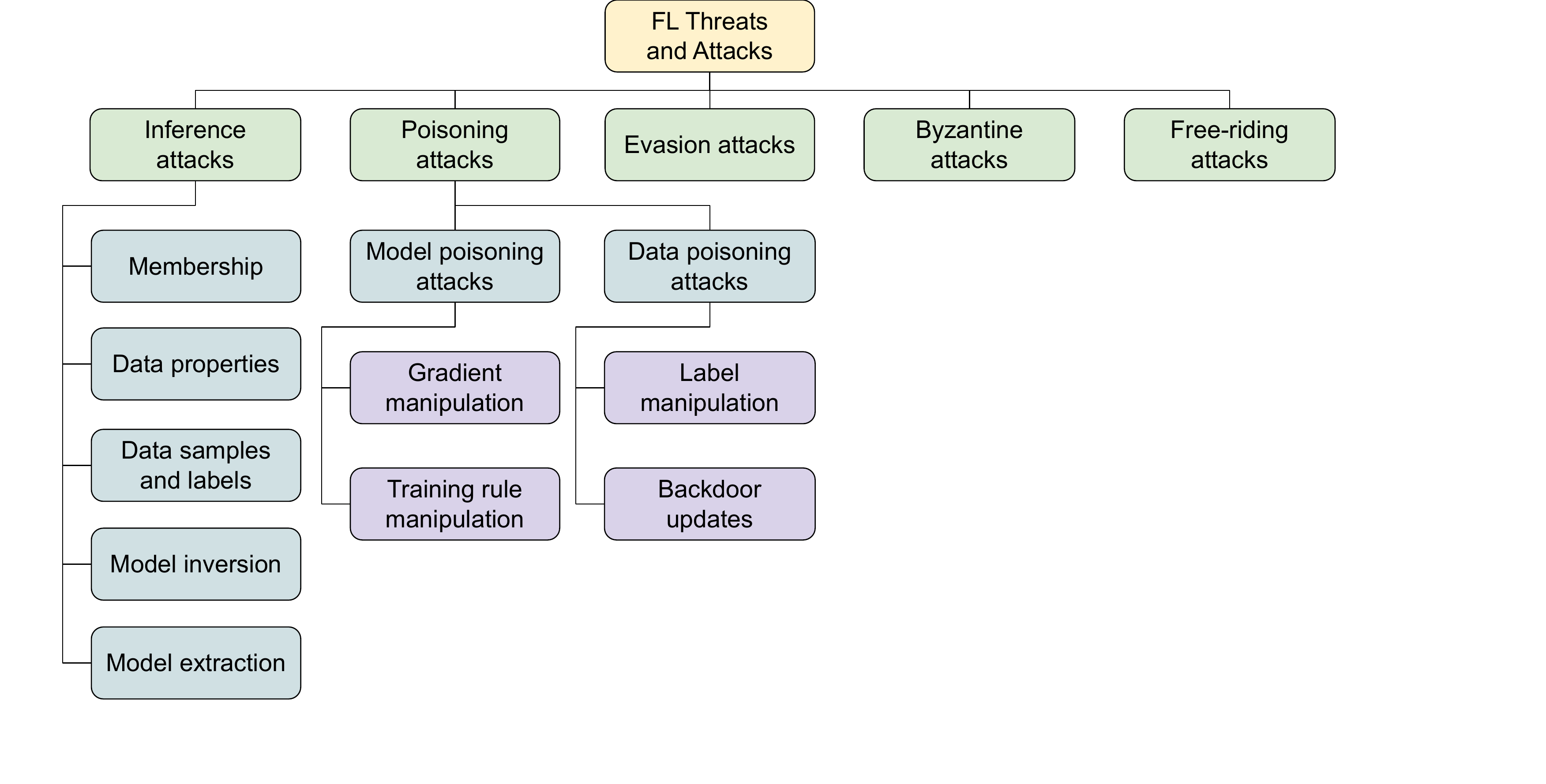}
    \caption{Classifying FL threats and attacks.}
    \label{fig:FL_attacks_tree}
\end{figure}

\subsubsection{Inference attacks}
\hfill\\
During inference attacks, an adversarial agent attempts to capture sensitive information throughout training. Training data, participant data, and label data are recorded without permission in hopes of correlating encrypted information with real values. Generative Adversarial Networks (GANs) can create powerful inference attacks. Inference attacks fall into five general categories:
\begin{enumerate}[label=(\roman*)]
        \item Membership inference attack - Attackers determine if an individual was present in the training dataset. The authors in \cite{shokri2017membership} constructed a membership inference attack that exploits the observation that machine learning models often behave differently on the data they were trained on versus the data they see for the first time  \cite{shokri2017membership}. Adversaries attempt to infer from specific samples that can be captured from model outputs. The objective is for adversaries to determine if a specific record exists in the model training dataset. 
    	\item Data properties inference attack - Data properties of a machine learning model are inferred using the parameters shared during model training. The particular data property is fully investigated to learn the frequency of the property within training data. For example, the authors in \cite{ganju2018property} demonstrated that a classifier that recognizes smiling faces also leaks information about the relative attractiveness of the individuals in its training set \cite{ganju2018property}. For example, a data properties inference attack on a model trained to screen FL participants by threshold may mistakenly reveal hardware or host types within the FL architecture. 
	    \item Data samples and labels inference attack - An adversary uses inference to capture targeted data samples and labels. FL model classes and participant training input labels can be reconstructed for inference. A malicious participant calculates distributed gradients to infer private information about other participants or model parameters. The authors in \cite{choquette2021label} demonstrated a label-only inference attack that could capture private information from machine learning models without access to confidence scores.   
	    \item Model inversion attack - Model inversion attacks aim to reconstruct private information from training data. Attackers with access to model parameters can look at the model's confidence score of predicted classes for inference. Successful model inversion attacks infer realistic representations from training data. The authors in \cite{fredrikson2015model} demonstrated the applicability of model inversion attacks on decision trees for lifestyle surveys as used on machine-learning-as-a-service systems and neural networks for facial recognition \cite{fredrikson2015model}. 
	    \item  Model extraction attacks - Adversary attack that aims to steal exact model parameters. Hu and Pang \cite{hu2021stealing}, mention how model extraction attacks aim to duplicate a machine learning model through query access to a target model. Hu and Pang studied two attacks on GANs: fidelity extraction attacks and accuracy extraction attacks \cite{hu2021stealing}.
        A strategic position can be extracted from learned functionality. Transfer learning attacks can occur when attacks are trained on vulnerabilities of similar models with the same framework, then transferred to attack the target model.  Extracting knowledge of the target models' training data and functionality increases future attack effectiveness. 
\end{enumerate}

\subsubsection{Poisoning attacks} 
\hfill\\
Poisoning attacks include misleading data injected by adversaries as inputs. Poisoning attacks can be classified into two categories: (1) model poisoning and (2) data poisoning attacks.
\begin{enumerate}[label=(\roman*)]
	\item Model poisoning attacks - In model poisoning attacks, the attacker reduces the model's performance on targeted sub-tasks (e.g., classifying planes as birds) by uploading "poisoned" updates \cite{panda2022sparsefed}. Poisoning attacks change the model's weights and biases, leading to misclassifying data. The model's confidence decreases when interacting with poisoned data. 
\begin{itemize}
	    \item Gradient manipulation attack - Transmitted gradients and parameters are manipulated throughout the FL training process to reduce global model performance. As a result, global model accuracy is reduced by adversaries injecting malicious updates. In addition, misclassification can result from gradient manipulation, allowing adversaries to increase the success rate of their attacks.
	    \item Training rule manipulation attack - An adversary attempts to minimize the difference between correct and incorrect training updates during training. The strength of this attack increases when multiple participants become dishonest and poison their parameters. Additionally, adversaries may bribe or collude with multiple participants to increase misclassification and lower detection mechanisms. The objective is for eventual harmful updates to go undetected based on the similarity of correct and incorrect training updates.
\end{itemize}
	\item Data poisoning attacks - Training data is altered to impact the FL model negatively. During training, contaminated data is introduced to corrupt the central aggregator. Data poisoning attacks inject malicious data into the training dataset before the learning process starts \cite{duan2022combined}. Such an attack allows adversaries to disguise harmful data, and eventually, the large amounts of poisoned data increase training times.
    \begin{itemize}
	    \item Label manipulation attack - Label manipulation attacks cause the model to mislabel training data. Label flipping includes training sample labels being flipped around to lower the judgment of the model. Classification errors can result from label flipping. Label flipping is an example of dirty-label attacks. In contrast, clean-label attacks are created when an adversarial makes poisoned training data indistinguishable from non-poisoned training data. Clean-label attacks are difficult but extremely powerful. 
	    \item Backdoor attack -  A backdoor attack tricks the model into associating a backdoor pattern with a specific target label so that, whenever this pattern appears, the model predicts the target label, otherwise, behaves normally \cite{liu2020reflection}. Backdoor attacks are also referred to as Trojan attacks. Adversaries inject clean or dirty backdoor updates into data samples in the training phase. When the global model performs aggregation, the model performance on the specific input is reduced depending on the number of undetected backdoors. When the backdoor is triggered, incorrect predictions occur on the targeted task while the main task performance appears untampered. Backdoor attacks compromise a subset of samples exchanged with adversaries during local training.
\end{itemize}
\end{enumerate}     
\subsubsection{Evasion attacks} 
\hfill\\
Evasion attacks - An adversary may attempt to evade a deployed system at test time by carefully manipulating attack samples \cite{biggio2013evasion}. Evasion attacks can bypass detection during training time with carefully manipulated attack samples. The input looks unaltered to humans but deceiving to the machine learning model. Evasion attacks contaminate training data during aggregation with undetectable misleading data. 
\subsubsection{Byzantine attacks} 
\hfill\\
Byzantine attacks - In a Byzantine attack, a malicious device gets processed with honest devices. Byzantine attacks aim to harm consensus and decrease model performance. 
In FL, a malicious attacker may control multiple clients, known as Byzantine users \cite{wen2022survey}. Byzantine users can upload fake data due to unreliable communication channels, corrupted hardware, or malicious attacks. This leads to the global model being manipulated by attackers and cannot be converged \cite{wen2022survey}.
    Larger amounts of 
    Byzantine users increase the effectiveness of Byzantine attacks. In the worst case, a 51\% attack could occur when the majority of participants are dishonest. 
\subsubsection{Free-riding attacks} 
\hfill\\
Free-riding attacks - Within FL environments, the global model is distributed to all participants regardless of their contributions. 
``Free-riders'' may emerge that do not contribute their fair share during the training process. Free-riders create free-rider attacks that lower model performance due to fake model updates being exchanged. A dishonest participant may become a free rider to withhold private information, reduce resource costs, or receive participation rewards while lacking participation requirements. The authors in \cite{fraboni2021free} mention two types of free-riders. Plain free-riders, which do not update the local parameters during the iterative federated optimization, and disguised free-riders, which employ sophisticated disguising techniques relying on stochastic perturbations of
the parameters \cite{fraboni2021free}.

\begin{table}[htp!]
\caption{Summary of FL attacks and resource costs.}
\tiny
\setlength\tabcolsep{6pt}
\begin{tabular}{|l|l|l|l|l|l|l|l|}
\hline  Attack type & Ref.& Interop. & Privacy risks &  \multicolumn{2}{l|}{Resource Consumption}  & Prevention & Impact \\
\cline{5-6} & & & & Adversary & Server & & \\
\hline Inference & \cite{shokri2017membership} & High & High & High & Medium & Medium & High \\
\hline Poisoning & \cite{panda2022sparsefed} & High & Medium & High & High & Medium & High \\
\hline Evasion & \cite{biggio2013evasion} & High & Low & High & Low & Low & Medium \\
\hline Byzantine & \cite{xia2020defenses} & Medium & High & High & High & Medium & High \\
\hline Free-riding & \cite{fraboni2021free} & Low & Low & Low & Medium & High & Medium \\
\hline
\end{tabular}
\end{table}

\subsection{Threats during FL phases}
\begin{figure}[htp]
    \centering
    \includegraphics[width=1\linewidth]{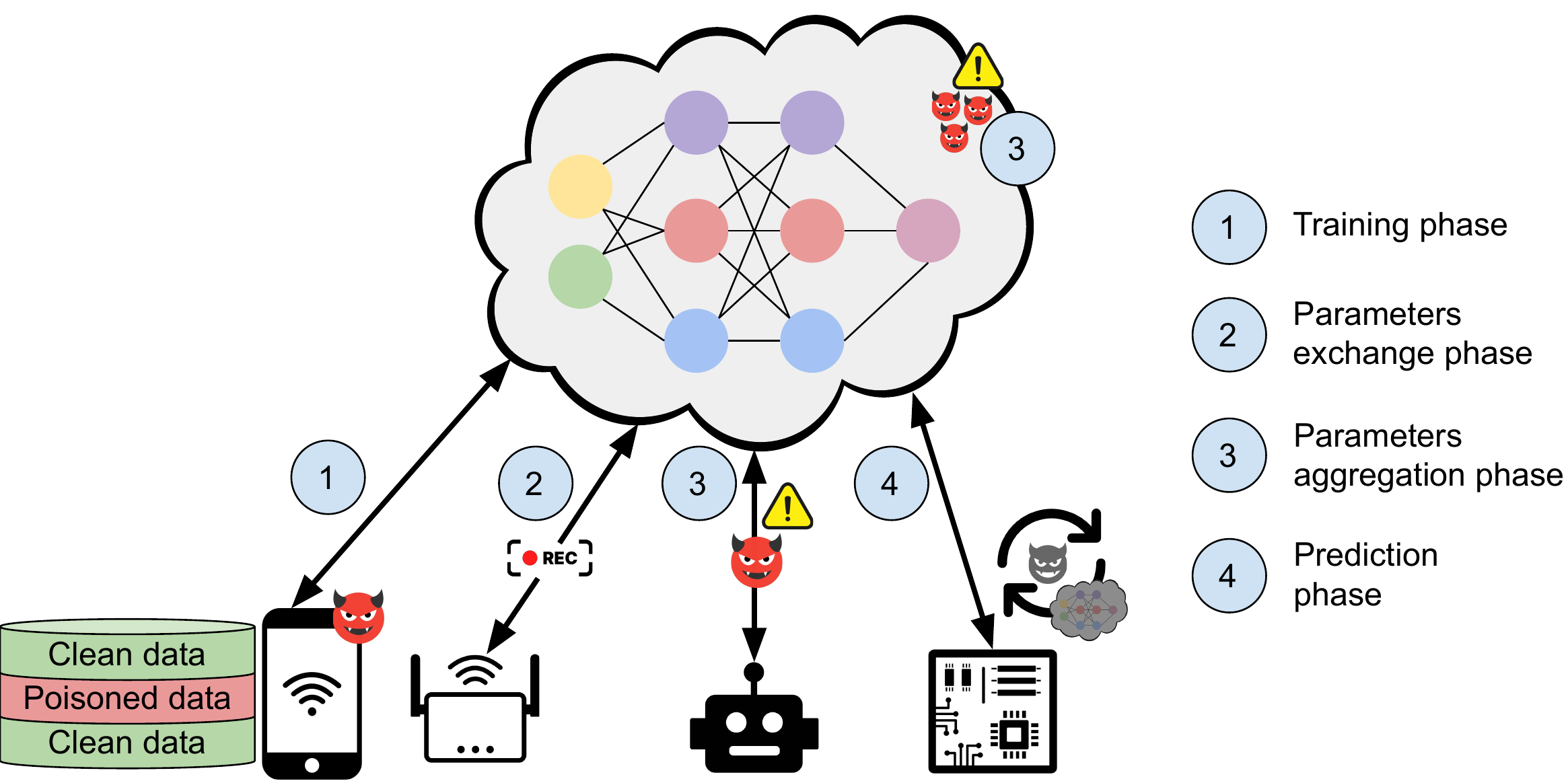}
    \caption{Potential threat observations within the four  phases of federated learning.}
    \label{fig:Threats_during_FLphases}
\end{figure}

A range of various attacks can compromise FL security. FL attacks can occur during four distinct FL phases: the training phase, parameters exchange phase, parameters aggregation phase, and prediction phase. Each phase has a different attack surface, most susceptible to model poisoning and inference attacks. While FL performs continuous learning, each phase is repeated in multiple iterations. As a result, undetected attacks strengthen as resources are drained from the environment. Four distinct FL phases and potential threats:
\begin{enumerate}[label=(\roman*)]
    \item Training phase - During the training phase, training data is sent to participants for learning. Local training data is not validated during this phase, allowing adversarial attacks. Early attacks in the training phase occur when the participant's reputation is unavailable. Malicious participants can pollute the training phase by conducting poisoning attacks. Adversaries may target the model performance for misclassification or overall training times through misleading data. Increasing amounts of adversarial data negatively impact model effectiveness. 
    \item Parameters exchange phase - The parameter exchange phase consists of A) participants downloading the global model parameters or B) participants uploading the local model parameters. External adversaries may attempt to capture sensitive information during parameter exchange through inference attacks. Inference attacks during this phase can steal parameters (model extraction), combine data to reconstruct the dataset, and estimate the distribution of continuously exchanged data. In addition, eavesdropping attacks can occur when communications are intercepted and potentially compromised online.
    \item Parameters aggregation phase - During the parameters aggregation phase, the central server performs a weighted average of participant parameters. Misleading parameters during this phase affect system-wide performance. If the central model architecture is well understood, adversaries can target the central server directly. The central server coordinates the aggregation of online learning and may learn to become malicious due to negative influences such as poisoning attacks. Fig. \ref{fig:Threats_during_FLphases} showcases the danger of adversarial updates being successfully processed into the parameter aggregation phase, thus lowering central server performance. Preserving central server effectiveness is important for the health of the FL architecture. Honest-but-curious central servers attempt to learn all possible information, overstepping boundaries through conducting inference or model poisoning attacks. Inefficient central servers incorrectly redistribute information to participants.
    \item Prediction phase - The prediction phase occurs after the global model has been deployed to devices. Adversaries may continually obtain the deployed global model to perform a combination of evasion and inference attacks. Evasion attacks deceive end devices, causing predictions to be inaccurate. Inference attacks on deployed predictions attempt to extract private information such as participant data, model parameter data, and training data.
\end{enumerate}

\subsection{Frog-boiling attacks in online environments}
Frog-boiling attacks reveal the limitations of anomaly detection in online environments. The authors in \cite{chan2011frog} proposed the frog-boiling attack, where an adversary disrupts the network while consistently operating within the threshold of rejection \cite{chan2011frog}. Frog-boiling refers to the phenomenon that a frog placed in hot water will instantly jump out. In contrast, a frog placed in slightly warm water that gradually heats up will remain in the water and eventually boil to death. Adversaries in online environments may continually attempt smaller attacks to learn the system's state and eventually disrupt the network. The authors in \cite{chan2011frog} mention three variants to the frog-boiling attack: the basic-targeted attack, the network-partition attack, and the closest-node attack \cite{chan2011frog}. Frog-boiling attacks can avoid outlier detection and increase network-wide latency.
\subsubsection{Insider and outsider attacks}
\hfill\\
FL environments can be threatened by the insider and outsider attacks. Insider attacks can be launched by either
the FL server or the participants in the FL system \cite{lyu2022privacy}. Insider attacks affect global model aggregation performance and can infer real values from noisy data. On the inside, a dishonest participant may attempt to decode FL updates through malicious tactics. Honest participants on the inside may be impersonated or bribed by external adversaries. In comparison, outsider attacks do not have direct access to internal communications of the network. From the outside, adversaries may eavesdrop on communications between the client and the server. Transferring data wirelessly can be intercepted, leading to compromised FL updates.
\subsubsection{Device latency}
\hfill\\
IoT devices are heterogeneous and fluctuate in resource constraints. Each device may have different amounts of noise depending on the environment. An FL central server waiting for participants to complete the training phase can be delayed by slower devices. Devices may have unreliable memory bandwidth or unsatisfactory hardware. Delays at the device level can cause delays throughout a synchronous FL architecture. According to Issa et al. \cite{issa2022blockchain}, the speed of rounds in synchronous FL is restricted to the speed of the slowest device. This causes a 
``straggler effect'', which can cause inefficient processing \cite{issa2022blockchain}.
\subsubsection{Consensus and dishonesty}
\hfill\\
Consensus performance is affected by the number of dishonest participants. The authors in \cite{aggarwal2021attacks} stated that the blockchain network security level is directly proportional to the amount of hash computing power that supports the blockchain. As the miners increase in the mining process, it becomes more difficult for an attacker to attack the blockchain \cite{aggarwal2021attacks}. The amount of honest and dishonest participants directly influences consensus mechanisms in FL and blockchain. Adversaries may employ Sybil attacks, where multiple fake identities are created to throw off consensus mechanisms. The weight of dishonest participants can cause consensus to malfunction. In the worst case, a 51\% attack can control network consensus. 
\subsubsection{Additional noise against adversaries}
\hfill\\
Additional noise creates a trade-off between levels of security and central server performance. 
The authors in \cite{blanco2021achieving} proposed that Federated learning intrinsically protects the data stored on each device by sharing model updates, e.g., gradient information, instead of the original data. However, model updates, which are based on original data, can reveal sensitive information \cite{blanco2021achieving}. Adding additional noise can increase the uncertainty of adversaries and reduce adversarial attack effectiveness. Jia et al. \cite{jia2019memguard} proposed to add a carefully crafted noise vector to a confidence score vector to turn it into an adversarial example that misleads the attacker's classifier \cite{jia2019memguard}.

\begin{table}[htp!]
\caption{\label{tab:table-name2} Expanded table of FL attacks and resource costs [Inter. = Interoperability, Ref. = Reference].}
\tiny
\setlength\tabcolsep{5pt}
\begin{tabular}{|l|l|l|l|l|l|l|}
\hline  Attack type & Interop. & Privacy risks &  \multicolumn{2}{l|}{Resource Consumption}  & Prevention & Impact \\
\cline{4-5} & & & Adversary & Server & & \\
\hline Inference & High & High & High & Medium & Medium & High \\
\hline Membership & High & High & High & Low & Medium & High \\
\hline Data properties & High & High & High & Low & Medium & High\\
\hline Data samples and labels & High & High & High & Low & Medium & High\\
\hline Model inversion & High & High & High & Low & Low & Medium\\
\hline Model extraction & Medium & Low & High & Medium & Medium & High\\
\hline Poisoning & High & Medium & High & High & Medium & High \\
\hline Model poisoning & High & Low & High & High & Medium & High\\
\hline Gradient manipulation & High & Low & High & Medium & Low & Medium \\
\hline Training rule manipulation & High & Low & High & Medium & Low & Medium \\
\hline Data poisoning & High & Low & High & High & Medium & High \\
\hline Label manipulation & High & Medium & High & Medium & Medium & Medium \\
\hline Backdoor & Medium & Medium & High & Medium & Medium & Medium\\
\hline Evasion & High & Low & High & Low & Low & Medium \\
\hline Byzantine & Medium & High & High & High & Medium & High \\
\hline Free-riding & Low & Low & Low & Medium & High & Medium \\
\hline
\end{tabular} \label{table3}
\end{table}

Table \ref{table3} contains details on how columns in the expanded table of FL attacks and costs were evaluated and scored. Interoperability was evaluated as the compatibility of attacks in an FL environment. In most cases, adversaries use high amounts of resources to conduct attacks. Some attacks explicitly drain the resources of the central server, while other attacks avoid server resource consumption to evade detection. The prevention column compares detection levels and proposed defenses for specified attacks. The impact column considers the importance of privacy preservation and architecture robustness. Well-planned combinations of attacks increase attack effectiveness.

\subsection{Vulnerabilities of FL and BCFL}
The attack surface of FL can be reduced by combining architectures with blockchains. FL has a large attack surface due to the single point of failure component of the FL system. An adversary could directly attack the central server to cause instability throughout the FL network. System-wide latency occurs when communications between devices and the central server are unreliable. Blockchain FL offers devices with distributed reliability for computing performance. When a device signals availability, the application layer of the blockchain automates a smart contract for a conditional agreement. Smart contracts on blockchains are immutable, so once created, the smart contract cannot be modified. Imperfect smart contracts include unwanted loopholes or vagueness.

Distributed machine learning architectures such as BCFL are vulnerable to outsider attacks while communicating over internet connections. Communication attacks such as eavesdropping attacks, man-in-the-middle attacks, and spoofing attacks are potential threats to the exchange of private information. A denial of service on network communications can be highly disruptive. Flooding a network with malicious data is resource-draining for distributed computing architectures.

Adversaries can commit poisoning attacks that reduce performance by sending deceptive inputs to mislead calculations. Dishonest participants redirect optimal resource allocations by providing inaccurate data. Byzantine users who upload fake data can purposely reduce global model performance. A 51\%
attack occurs when the number of dishonest participants outweighs the number of honest participants; thus, most dishonest participants can gain control of an unbalanced network.

\begin{figure}[htp!]
    \centering
    \includegraphics[width=\linewidth]{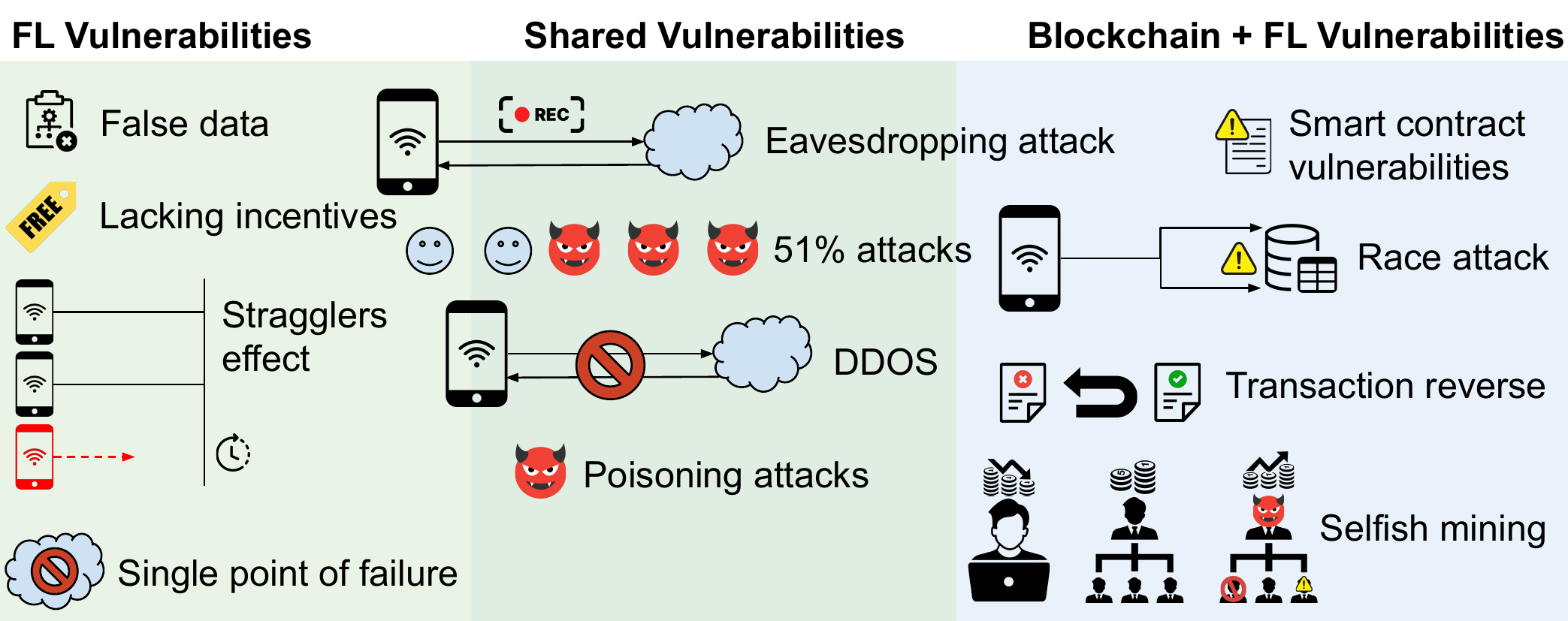}
    \caption{Various vulnerabilities found in FL and BCFL.}
    \label{fig:fig_vulnerabilites_FL_vs_BCFL}
\end{figure}

The authors in \cite{yang2019effective} mention the cost of creating a 51\% attack is surprisingly low if hash power is abundantly available \cite{yang2019effective}. In FL, a 51\% attack can mislead FL training when the majority of participant data is untrustworthy. In blockchain, a 51\% attack can control the consensus mechanisms that intervene with transaction authentication. Blockchain transaction authentication can be delayed when conflicting transactions are completed at the same time. A race attack creates a potential fork in the blockchain within the short window of time required to authenticate transactions. Selfish mining pools may purposely generate or withhold blocks simultaneously to gain an advantage over other miners. Transaction reverse is a dishonest mining tactic where the processing of potentially generated blocks is reversed to cause Blockchain delays.

\subsection{Integration Motivation of FL and Blockchain}
Blockchains have inherent interoperability, allowing blockchains to communicate with other blockchains efficiently. Blockchain interoperability increases robustness when useful information is transferred between blockchains. For example, if a blockchain layer undergoes a denial of service attack, a similar blockchain could offer alternative services to supplement security. Similarly, participant data such as reputation can be communicated between blockchains to increase scalability. Blockchain can solve the problem of trust establishment among distributed systems through distributed node verification, and consensus mechanisms \cite{li2021blockchain} \cite{yang2018blockchain}.

Optimal device selection reduces the dangers of adversaries. A device selection phase can reduce resource costs associated with unfavorable behavior. Ideally, the behavioral patterns between an honest and dishonest participant are largely different. Behavioral auditing of malicious patterns from questionable participants can increase security. The authors in \cite{barreno2010security} proposed the Reject On Negative Impact (RONI) defense to determine whether a candidate training instance is malicious. \cite{barreno2010security} states if adding the candidate instance to a training set causes the resulting classifier to produce substantially more classification errors, reject the instance as detrimental in its effect.

\section{Learning on Resource-Constrained Devices}
Latency on IoT devices can increase based on the distance between data centers. Solutions to device latency include moving cloud resources closer to devices, which is seen in Mobile edge computing (MEC). The authors in \cite{nguyen2021federated} determined that MEC servers are now becoming a weak point due to data privacy concerns and high data communication overheads. Mobile devices may be subject to attack if not properly secured, and may require additional security when deploying important communications in MEC. The authors in \cite{li2021blockchain} mention how blockchain can solve the security problem of edge computing. 

As we know, IoT devices may have limited resources, and quantifying the expected resource usage can help reduce excess resource expenses \cite{imteaj2019distributed}. If a device is showing signs of being a straggler, device dropout can be applied to reduce communication costs. However, if we have a large number of stragglers, then simply dropping them can put a negative impact on the overall learning process. The authors in \cite{imteaj2020fedar} developed FedAR algorithm that can select only the effective and trustworthy agents for learning process. Besides, FedPARL framework is specially designed for resource-constrained FL devices that can prune large model and allow feasible local task allocation for the edge devices \cite{imteaj2021fedparl}. To deal with heterogeneous resources of the agents, the authors in \cite{9680096} proposed an FL model that can generate multiple global models considering the resource status of the agents and accelerate the learning process. Moving forward, increasing device participation can cause diminishing returns if devices

\begin{figure}[htp!]
\begin{center}
    \includegraphics[width=.8\linewidth]{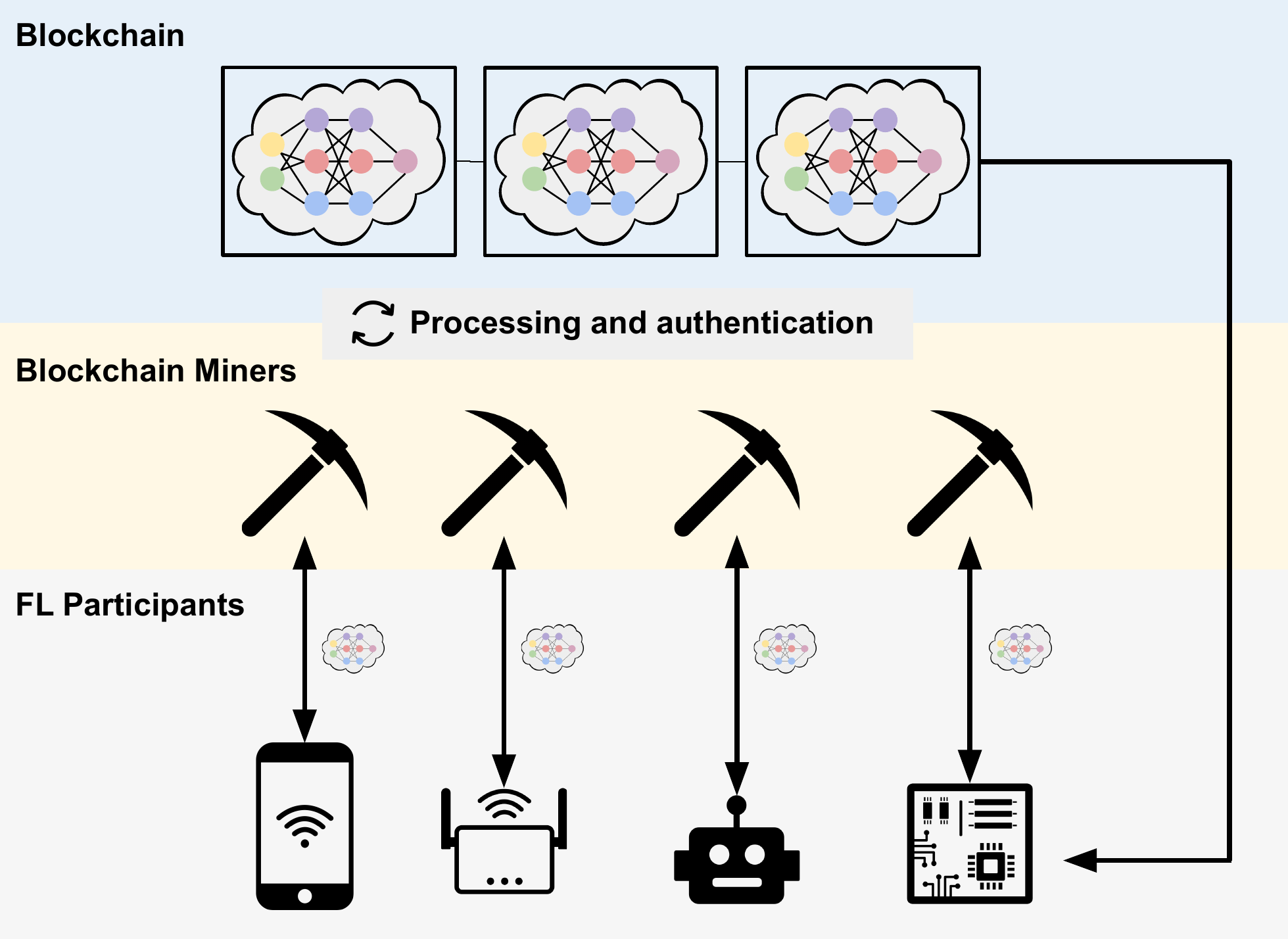}
    \caption{Condensed workflow with integrated blockchain-based FL.}
    \label{fig:BCFL_architecture}
\end{center}
\end{figure}

\noindent perform poorly during training. Adaptive methodologies, such as adaptive FL can help balance resource costs. The authors in \cite{wang2019adaptive} proposed an adaptive FL control algorithm based on gradient-descent for reducing computational resource budgets. The proposed control algorithm determined the best trade-off between local update and global parameter aggregation to minimize the loss function under a given resource budget \cite{wang2019adaptive}. The authors in \cite{baccarelli2022afafed} proposed AFAFed, an Adaptive FL solution for IoT devices affected by packet-loss communication. One of the key features of AFAFed is the implementation of distributed sets of local adaptive tolerance thresholds and global centralized adaptive fairness coefficients. AFAFed's features allow the algorithm to calculate the right personalization vs. fairness trade-off in various resource-constrained computing environments.

\subsection{Verification of Local Model Updates}
Verifying local model updates helps prevent resource-consuming\footnote{Resource-consuming can be considered negative, harmful, or malicious data.}
data from being aggregated with the global model. The authors in \cite{wei2020federated} found that blockchain-assisted decentralized FL frameworks can prevent malicious clients from poisoning the learning process and thus provides a self-motivated and reliable learning environment for clients \cite{wei2020federated}. Blockchain consensus mechanisms boost network agreement, thus deterring unreasonable data from being aggregated. The authors in \cite{yang2022trustworthy} mentions four different blockchain consensus protocols for confirming the correctness of a global model and executing global model aggregation. Blockchain consensus protocols include: Proof of Work (PoW), Proof of Stake (PoS), Raft, and practical Byzantine fault tolerance \cite{yang2022trustworthy}.

\begin{itemize}
    \item Proof of Work (PoW) discourages harmful blocks from being added to the network by requiring the completion of a cryptographic puzzle by miners before verification. The authors in \cite{lee2022dag} proposed that PoW is the most famous algorithm
for Bitcoin. Although it works effectively for protection from Sybil attacks and data
manipulation, it is expensive because of the required hash power and long block interval  \cite{lee2022dag}.
    \item Proof of Stake (PoS) randomly selects data validators based on a total stake or overall impact on the network. Individuals with high amounts of holdings are given the opportunity to validate and approve new blocks being added. Validators in PoS create validator nodes to assist in network consensus. The validator's assets are locked while staking to help support network consensus. Validators receive incentives for vouching for the legitimacy of transactions.
    \item Raft supports consensus within decentralized blockchain environments. A raft can improve the understandability of the verification by breaking consensus problems down using decomposition. According to Ongaro and Ousterhout \cite{ongaro2014search},  Raft separates the key elements of consensus, such as leader election, log replication, and safety, and it enforces a stronger degree of coherency to reduce the number of states that must be considered \cite{ongaro2014search}.
    \item Practical Byzantine fault tolerance (PBFT) ensures a decentralized network continues operating even if a portion of nodes fail or act maliciously. The authors in \cite{zhang2020qpbft} mention that PBFT is an optional consensus protocol for consortium blockchains scenarios, where strong consistency is required \cite{zhang2020qpbft}. Consortium BCFL can prevent the straggler effect in FL, where the speed of the slowest device causes network-wide delays.

\end{itemize}

\subsection{Global Model Aggregation}
Global model aggregation continually occurs online. Blockchains can provide decentralized applications, such as smart contracts, to achieve global model aggregation without a central server. The distributed training of the global model can encourage dishonest participants to collude with others and launch a coordinated attack. The authors in \cite{bouacida2021vulnerabilities} suggested that past clients can coordinate with current or future participants to participate in attacks against current or future updates to the global model \cite{bouacida2021vulnerabilities}. Global model updates may not change much depending on context. While global model parameters are not changing much, parameters can be frozen to reduce communication costs. The authors in \cite{chen2021communication} found that it is unnecessary to always synchronize the full FL model in the entire training process because many parameters gradually stabilize prior to the ultimate model convergence \cite{chen2021communication}. Devices with low bandwidth benefit from frozen parameters to reduce device resource costs. Yang et al. \cite{yang2022trustworthy} proposes a decentralized blockchain-based FL architecture that can resist failures or attacks of servers and devices by building trustworthy global model aggregation with secure model aggregation based on blockchain consensus protocol among multiple servers.
\subsection{Incentive Mechanism}
Federated learning does not reward local device participation. Blockchain can offer an incentive mechanism to improve participation. The authors in \cite{qu2022blockchain} mention that there could be heterogeneous devices with different computational and data resources in an FL system. Therefore, participants with better resources must have extra benefits compared to participants with little contribution \cite{qu2022blockchain}. Honest contribution is advantageous to all participants and should be rewarded appropriately. Besides, the authors in \cite{zhu2022secure} mention how the blockchain's incentive mechanism can track the contribution of each data provider towards the globally optimized model so that participants can be treated fairly, thereby attracting more data sharers. Blockchain-based federated learning architectures such as BlockFL \cite{kim2019blockchained}, enable exchanging devices' local model updates while verifying and providing their corresponding rewards \cite{kim2019blockchained}. While the authors in \cite{ma2022federated} found that BCFL enables all clients to verify the learning results that are recorded on the blockchain, whereby distributed clients can be rewarded incentives to participate, and untrusted learning models can be detected.
\begin{figure}[htp]
    \centering
    \includegraphics[width=\linewidth]{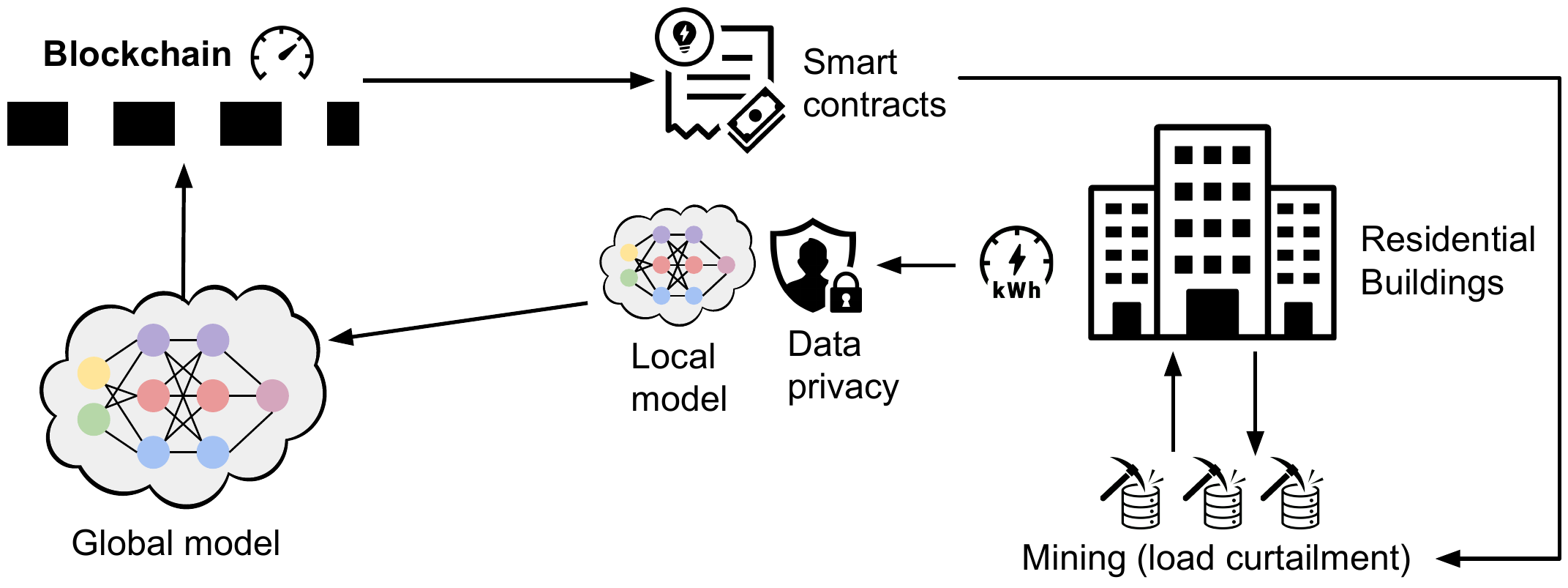}
    \caption{Example BCFL figure with example incentive mechanism for residential buildings.}
    \label{fig:vulnerabilites_FL_vs_BCFL}
\end{figure}

\subsection{Privacy and Security Protection}
Adding noise to data can reduce overall efficiency. Too much noise and performance are diminished, while not enough noise and the state is potentially insecure. Levels of noise carry a trade-off that should be considered. FL can ensure data privacy by only exchanging model parameters with devices. Local device data is never communicated online, boosting the security of data. Multiple approaches examine noise trade-off. Truex et al. \cite{truex2019hybrid} proposed an FL approach that utilizes both differential privacy and secure multiparty computation (SMC). \cite{truex2019hybrid} combined differential privacy with secure multiparty computation to enable the growth of noise injection to be reduced as the number of parties increases without sacrificing privacy. 

Security measures include anticipating anomalous behavior. Typical technologies for defending FL security include malicious participant detection, and malicious impact mitigation \cite{duan2022combined}. Irregular behavior can be indicative of an adversary or potential machine learning attack. The authors in \cite{hassan2022anomaly} mention six anomaly detection models for effectively detecting anomalous behavior: generative architectures, classification-based models, clustering-based models, nearest neighbor models, statistical \& analytical models, and reinforcement learning-based models \cite{hassan2022anomaly}. In addition, participant contribution can be measured to classify behavior between honest vs. dishonest participation.

\section{Existing Research Challenges and Future Research Directions}
\subsection{Potential Solutions of Existing Research Challenges}

Prior works demanded high resource consumption for large scale data-driven environments. FL and its form of distributed computing can process participation from different IoT devices, thus allowing a variety of resourceful participation. Resourceful participation has three main challenges:
\begin{enumerate}[label=(\roman*)]
\item High communication costs can be discouraged based on network reliability. Communication delays from service providers may require a device to drop out at slower speeds. The authors in \cite{cui2022fast} examine compressed communication to reduce communication overheads in BCFL. Cui et al., \cite{cui2022fast} created a communication-efficient framework that could reduce the training time by about 95\% without compromising model accuracy. Approaches that reduce communication overheads are advantageous for lowering participation requirements.
\item Trade-offs between security and performance exist. Increasing noise against participants boosts security but also lowers performance. Choosing the best devices for participation can reduce the dependence on trade-offs between security and performance. Consortium blockchains are present solutions for the possible diminishing returns of open participation. Consortium blockchains screen participating devices to encourage optimal participation selection of trusted devices. The authors in \cite{mohammed2020budgeted} propose a budgeted number of candidate clients chosen from the best candidate clients in terms of test accuracy to participate in the training process. Optimal selection strategies can verify participants are resourceful and honest before inviting these participants into training. 
\item Dishonest participation records are not shared between blockchains. Communicating participation records between blockchains can improve scalability. Records of dishonest behavior can warn similar blockchains of a participant's integrity.  For example a dishonest mining pool with evidence of performing mining attacks against a blockchain can be blacklisted. Blockchains interoperability can allow participants to be chosen thoughtfully.
\end{enumerate}
BCFL can improve resource-constrained environments, although being considerate of optimal participation further increases overall resource effectiveness. Li et al., \cite{li2022blockchain} suggests that blockchain can further improve FL security and performance, besides increasing its scope of applications.

\begin{figure}[htp]
    \centering
    \includegraphics[width=1.1\linewidth]{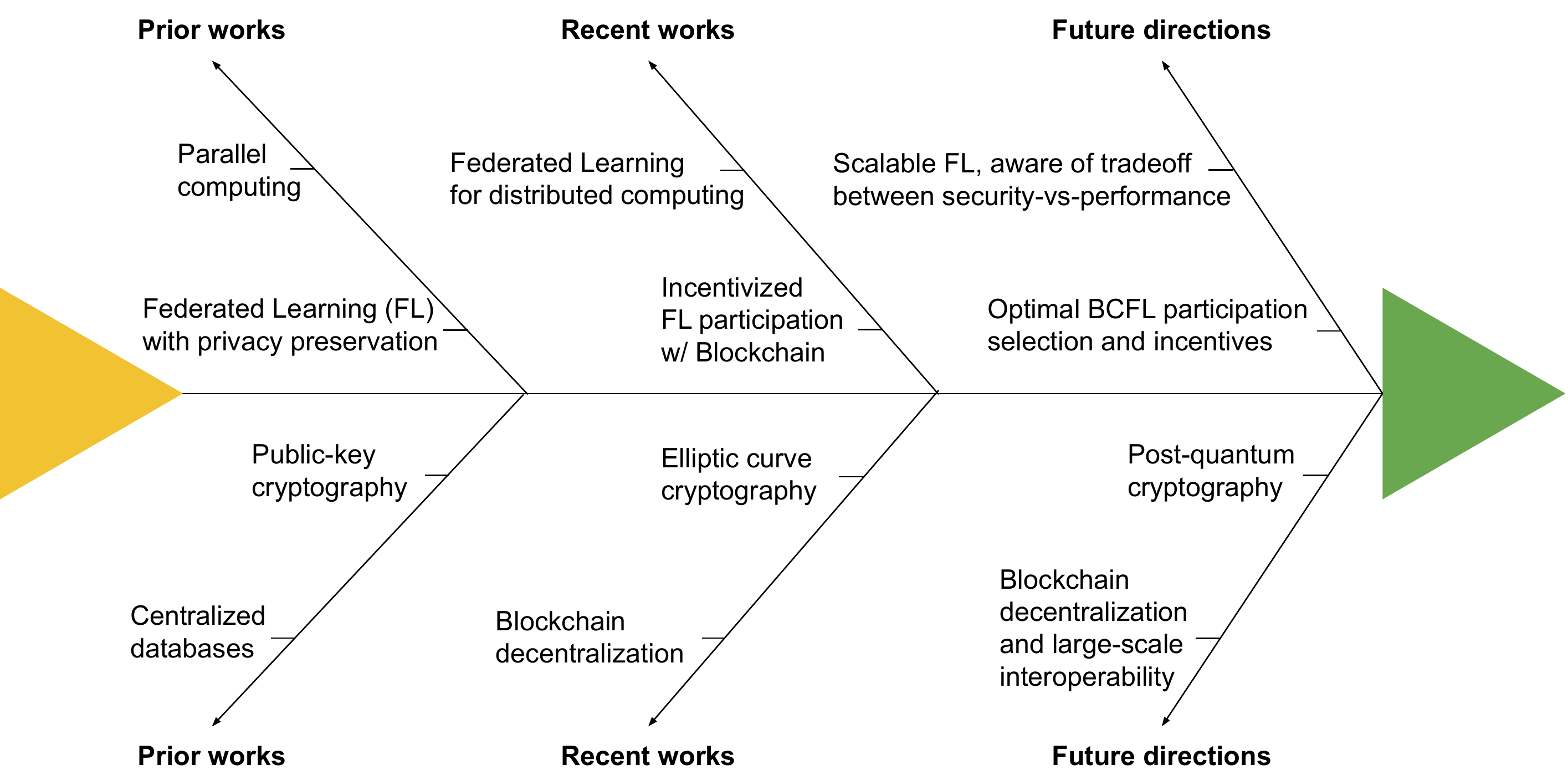}
    \caption{BCFL flow from prior works towards future directions.}
    \label{fig:BCFL_Fishbone}
\end{figure}

\subsection{Future Research Directions}
Future research can be conducted in the areas of BCFL scalability, quantum resilience and AI alignment. The aforementioned directions have challenges that may need to be addressed in the future. Figure \ref{fig:BCFL_Fishbone} displays a timeline of important BCFL research directions. Each direction can disrupt resource management, requiring revolutionary system designs:
\begin{itemize}
\item Scalability can cause diminishing returns when resource requirements exceed reasonable blockchain computing limits. The authors in \cite{imteaj2021foundations} states that as the number of transactions increases, it requires a larger blockchain size to store those transactions. However, mining a large size of blockchain may require more resources, which would be difficult for IoT devices \cite{imteaj2021foundations}. Dimensional reduction for unreasonably large blockchains can prevent future mining participants from requiring supercomputers, compared to current GPUs.

\item Post-quantum cryptography algorithms may be required for resilience from quantum attacks. The authors in \cite{fernandez2020towards} mention how the fast progress of quantum computing has opened the possibility of performing attacks based on Grover's and Shor's algorithms. Such algorithms threaten public-key cryptography, and hash functions \cite{fernandez2020towards}. Blockchain may require a cryptographic redesign to combat quantum attacks. Quantum computers bring unique threats to classical data management techniques.

\item AI alignment is important for building a favorable machine learning model and central server. Corrigibility, the capability of being reparable, can be considered when the machine learning model learns unfavorable behavior. The model may seize to optimize due to model convergence. Further research could evaluate a models curiosity and the relation to model convergence, when significant performance changes are relatively stable. An honest-but-curious central server may or may not be reparable. 
\end{itemize}

\section{Conclusion} 

In this paper, we present legible insights of blockchain-based FL from  crucial and time-demanding perspectives. Blockchain provides additional security in a decentralized format that protects FL from various real-world security and privacy issues. We categorize various threat models and point-out the leading vulnerabilities that could be observed during various stages of the FL process and in online environments of blockchain-enabled FL settings. To this end, we show a clear direction on how we can leverage secure and private learning on resource-constrained devices, covering the feasible verification of local model updates, secure global model aggregation, designing fair incentive scheme, and upgrading security and protection. Finally, we present the existing blockchain-based FL applications and highlight the potential solutions of the existing research challenges in the relevant domains. We anticipate that this survey will be helpful for researchers, practitioners, and scientists, developing robust blockchain-enabled FL systems choosing suitable consensus mechanism, identifying security and privacy pitfalls, motivating coherent formations, and following the promising future directions presented in this paper.

\section{{acknowledgment}}
This material is based upon Ervin Moore’s work supported by the U.S. Department of Homeland Security under Grant Award Number, 2017-ST-062-000002. The views and conclusions contained in this document are those of the authors and should not be interpreted as necessarily representing the official policies, either expressed or implied, of the U.S. Department of Homeland Security.

\ifCLASSOPTIONcaptionsoff
  \newpage
\fi



%

\bibliographystyle{unsrt}
\bibliography{main.bib}

\begin{IEEEbiography}[{\includegraphics[width=1in,height=1.25in,clip,keepaspectratio]{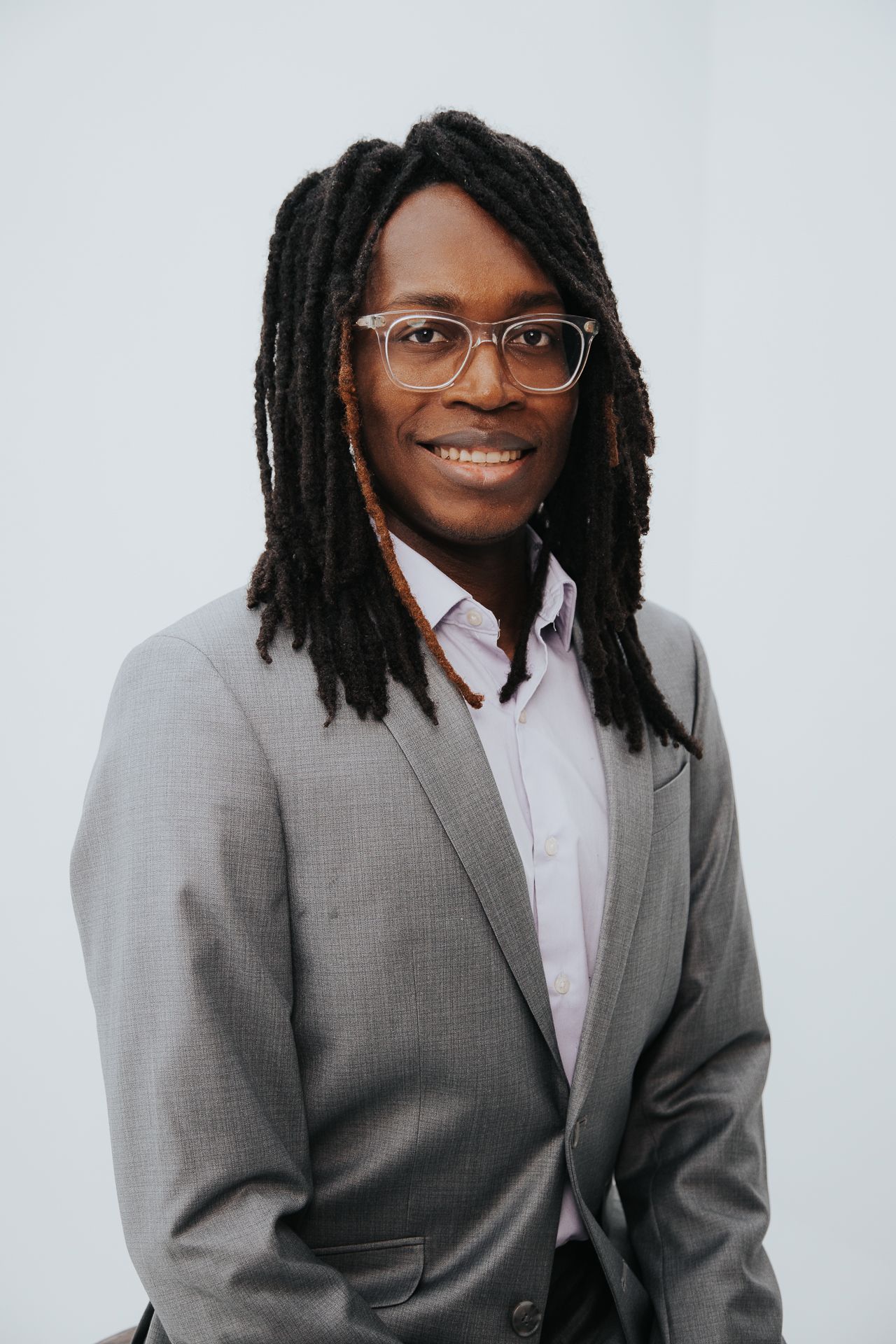}}]{Ervin Moore}  is a Ph.D. student and DHS CAESCIR Fellow at the Knight Foundation School of Computing and Information Sciences at FIU. He is a member of the Sustainability, Optimization, and Learning for InterDependent networks laboratory (solid lab) working with Dr. Amini and Dr. Rezapour. Prior to that, he graduated with a MSc in Artificial Intelligence and Machine Learning from Colorado State University-Global, following his BA in Communication Technology from University of Texas at Arlington. Ervin participated in AI-related hackathons and has won multiple. His applied research involved using machine learning and large language models to improve user experiences. His current research interests are artificial intelligence, machine learning and their application in critical infrastructure security and resilience.

\end{IEEEbiography}

\begin{IEEEbiography}
[{\includegraphics[width=1in,height=1.25in,clip,keepaspectratio]{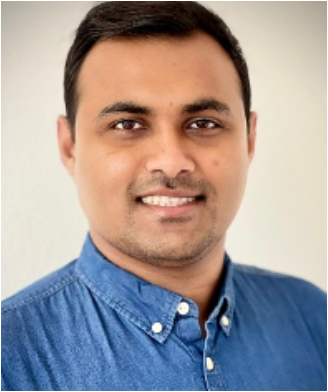}}]{Ahmed Imteaj} is an Assistant Professor in the School of Computing at Southern Illinois University, Carbondale. He received his Ph.D. in Computer Science from Florida International University in 2022, where he received his M.Sc. degree with recognition of the Outstanding Master’s Degree Graduate Award. Prior to that, he received B.Sc. degree in Computer Science and Engineering from Chittagong University Engineering and Technology. His research interests span Federated learning, cybersecurity, Internet of Things (IoT), and blockchain. 
Ahmed significantly contributed to the area of privacy-preserving distributed machine learning and internet-of-things and published his research in top-tier conferences and peer-reviewed journals. Ahmed's research has been recognized several times. Ahmed was recognized with the 2022 FIU Real Triumph Graduate Award, ``2022 Outstanding Student Life Award: the Graduate Scholar of the Year Award", ``2021 Best Graduate Student in Research Award" from the KFSCIS at FIU, Best Paper Award from the 2019 IEEE CSCI’19 conference, and one of the two winners at 2021 FlU GSAW Scholarly Forum. Ahmed is the lead author of the book, ``Foundations of Blockchain: Theory and Applications" and published more than 50 peer-reviewed journals and conference papers. 
\end{IEEEbiography}

\begin{IEEEbiography}[{\includegraphics[width=1in,height=1.25in,clip,keepaspectratio]{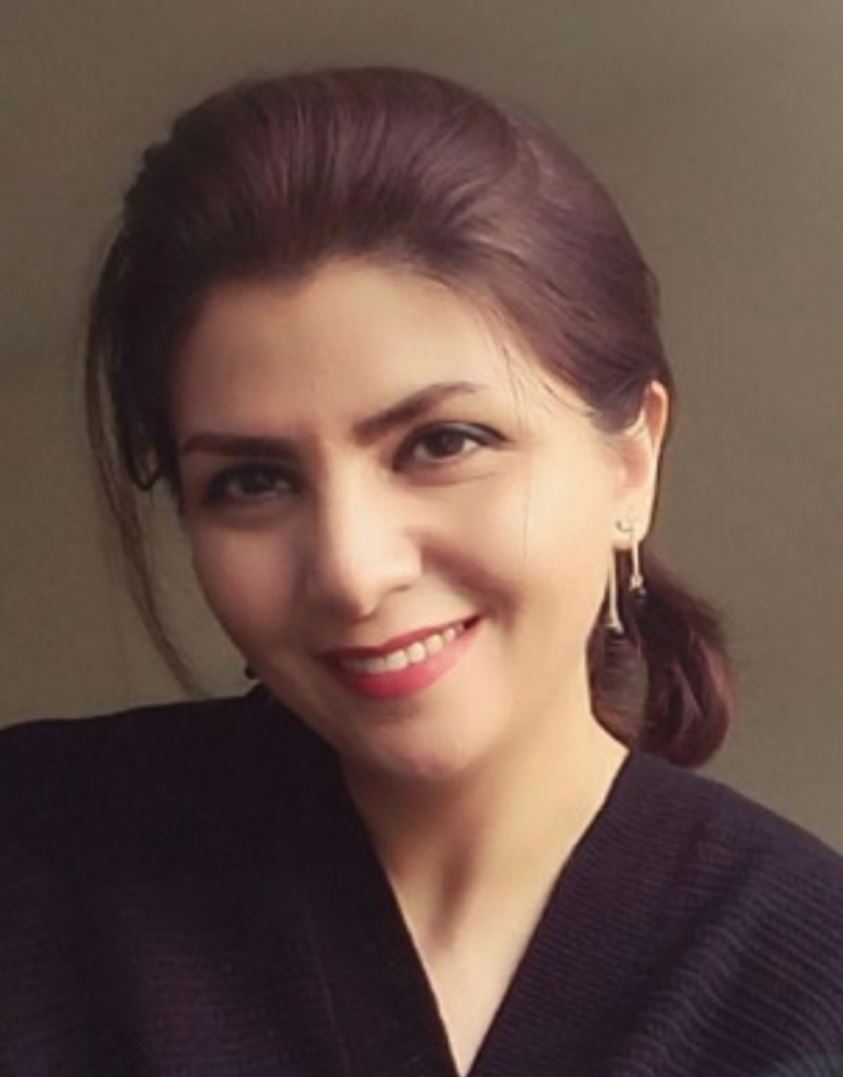}}]{Shabnam Rezapour}  is an Assistant Professor of Enterprise and Logistics Engineering at Florida International University. Her research focus is on “Architecting Resilient Societies” that are able to cope with and transform in the face of expected and unexpected chronic stresses (slow-moving disasters that weaken the fabric of societies such as climate change and water shortage) and acute shocks (sudden events that threaten societies such as floods, hurricanes, terroristic attacks, and disease outbreaks). Her research group develops analytical models (e.g., stochastic optimization and simulation models) to understand and analyze the vulnerability of network-centric infrastructures/systems and develop optimal mitigation policies. 
\end{IEEEbiography}

\begin{IEEEbiography}[{\includegraphics[width=1in,height=1.25in,clip,keepaspectratio]{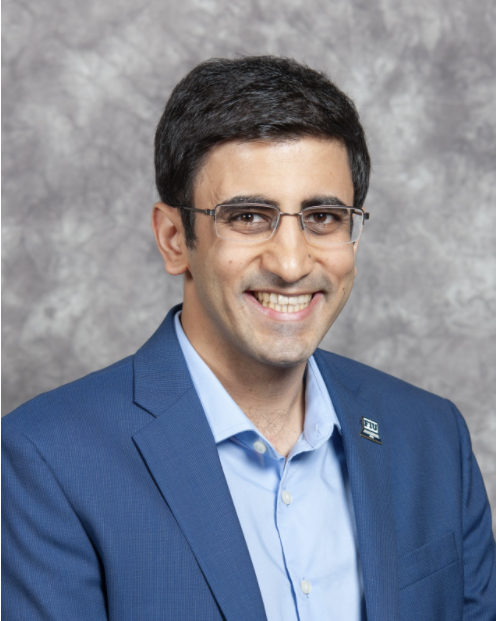}}]{M. Hadi Amini} (S’11–M’19-SM'22) is an Assistant Professor at Knight Foundation School of Computing and Information Sciences at Florida International University. He is the director of Sustainability, Optimization, and Learning for InterDependent networks laboratory (www.solidlab.network). He received his Ph.D. in Electrical and Computer Engineering from Carnegie Mellon University in 2019, where he received his M.Sc. degree in 2015. He also holds a doctoral degree in Computer Science and Technology. Prior to that, he received M.Sc. degree from Tarbiat Modares University in 2013, and the B.Sc. degree from Sharif University of Technology in 2011. His research interests include distributed optimization and learning algorithms, distributed computing and intelligence, sensor networks, interdependent networks, and cyber-physical-social resilience. Application domains include smart cities, energy systems, transportation networks, and healthcare. He has received over \$3M in funding from various  federal and local funding agencies in the United States.

Hadi is a Senior Member of IEEE, and also a life member of IEEE-Eta Kappa Nu (IEEE-HKN), the honor society of IEEE. He served as President of Carnegie Mellon University Energy Science and Innovation Club; as technical program committee of several IEEE and ACM conferences; and as the lead editor for a book series on ``Sustainable Interdependent Networks'' since 2017. He also serves as Associate Editor of SN Operations Research Forum and International Transactions on Electrical Energy Systems. He has published more than 100 refereed journal and conference papers, and book chapters. He edited/authored six books. He is the recipient of the best paper award from ``2019 IEEE Conference on Computational Science \& Computational Intelligence'', FIU's Knight Foundation School of Computing and Information Sciences’ ``Excellence in Teaching Award'', best reviewer award from four IEEE Transactions, the best journal paper award in “Journal of Modern Power Systems and Clean Energy”, and the dean’s honorary award from the President of Sharif University of Technology. (Homepage: www.hadiamini.com)
\end{IEEEbiography}
\end{document}